\documentclass[a4paper,11pt]{article}
\pdfoutput=1 % if your are submitting a pdflatex (i.e. if you have
\usepackage{jheppub} % for details on the use of the package, please
\usepackage[T1]{fontenc} % if needed

\usepackage[english]{babel}

\def\ve {\varepsilon}

\def\b {\beta}
\def\g {\gamma} \def \G {\Gamma}
\def\d {\delta} 
 \def\L {\Lambda}
\def\m {\mu}

\def\p {\pi}

\def\dd{{\rm d}}

\def\({\left(}
\def\){\right)}
\def\[{\left[}
\def\]{\right]}
\def\Disc{\mathrm{Disc}}

\def\beq {\begin{equation}}
\def\eq {\end{equation}}
\def\bea{\begin{eqnarray}}
\def\eea{\end{eqnarray}}
\def\bf{\textbf}
\newcommand{\notd}[1] { \setbox0=\hbox{$#1$}
\dimen0=\wd0   \setbox1=\hbox{/} \dimen1=\wd1  \ifdim\dimen0>\dimen1
 \rlap{\hbox to \dimen0{\hfil/\hfil}}  #1 \else \rlap{\hbox to \dimen1{\hfil$#1$\hfil}}  /  \fi  }
\def\dd{{\rm d}}
\def\nn{\nonumber}

\title{\boldmath Two-loop massive scalar three-point function in a dispersive approach}

\author[a,b,1]{Vladyslav Pauk,\note{Corresponding author.}}
\author[a,b]{Marc Vanderhaeghen}

\affiliation[a]{Institut f\"ur Kernphysik, Johannes Gutenberg Universit\"at, Mainz D-55099, Germany}
\affiliation[b]{PRISMA Cluster of Excellence, Johannes Gutenberg-Universit\"at,  Mainz, Germany}

\emailAdd{paukvp@gmail.com}

\abstract{We present a dispersion relation formalism to calculate a massive scalar two-loop vertex function. 
Such calculation is of direct relevance in the evaluation of the hadronic light-by-light contribution to the muon's anomalous magnetic moment due to meson poles. 
The discontinuity of  the two-loop diagram is obtained by a sum of two- and three-particle cut contributions, 
which involve a phase space integration over the physical intermediate states. 
The real part of the vertex function is subsequently reconstructed through evaluation of a dispersion integral. We explicitly demonstrate that the dispersive formalism yields exactly the same result as the direct two-loop calculation. }

\begin{document} 
\maketitle
\begin{figure}[h]
\centering
  \includegraphics[width=4.5cm]{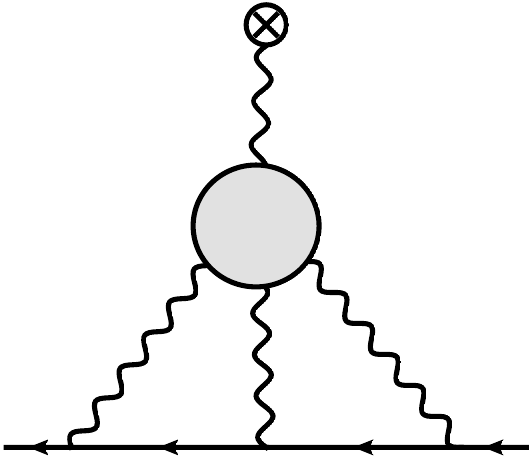}\\
  \caption{The hadronic light-by-light scattering contribution to the anomalous magnetic moment of the muon}
  \label{fig:LbL}
\end{figure}
\begin{figure}[h]
  \includegraphics[width=15cm]{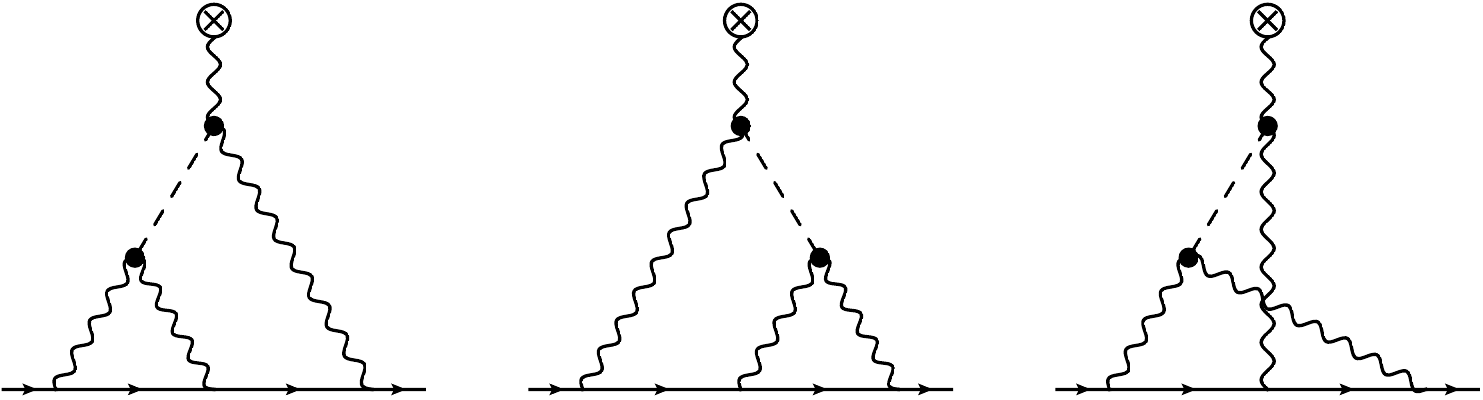}\\
  \caption{The meson-pole contributions to the hadronic light-by-light scattering in $(g-2)_\m$.}
  \label{LbL_pole_disp}
\end{figure}

\flushbottom
The computation of Feynman integrals is an essential part of the calculations within the Standard Model (SM) of particle physics. Dispersion relations provide a powerful tool for the calculation of such loop integrals in perturbation theory. They e.g. allow to reduce one-loop Feynman integrals to tree-level calculations followed by phase-space integration and evaluation of the dispersion integral. This procedure may be extended also to higher-order corrections (see, for instance Refs.~\cite{Schilcher:1980kr,Chang:1981qq,vanNeerven:1985xr,Kniehl:1989yc,Korner:1995xd} for the two-loop case, and Ref.~\cite{Kniehl:1989bb} for the three-loop case). Moreover, dispersion approaches offer unique ways to handle phenomena where perturbation theory is unreliable, particularly when the strong interactions are involved. 
In many cases, the dominant part of the uncertainty in the SM calculations originates from hadronic effects and the only reliable way to constrain them is based on experimental input. To this end, one exploits the unitarity and analytical properties of the non-perturbative amplitudes to relate them to the measurable correlation functions. 
A notorious example of such a situation is the evaluation of the hadronic light-by-light (HLbL) contribution to the anomalous magnetic moment of the muon (for a review, see \cite{Jegerlehner:2009ry} and references therein). The diagram appearing in this case is depicted in Fig. \ref{fig:LbL} and corresponds to a massive two-loop three-point function type. 
The singularities of the HLbL tensor contributing to the diagram correspond by unitarity to the production of physical hadronic states 
(entering the blob in Fig.\ref{fig:LbL}). 
The simplest type of diagrams appears when we consider the single meson exchange in the HLbL tensor and correspond with the diagrams in Fig. \ref{LbL_pole_disp}. 
The direct two-loop calculation of such diagrams for single meson exchange as in Fig.~\ref{LbL_pole_disp} involves the inclusion of meson 
form factors (entering the black blobs in Fig.~\ref{LbL_pole_disp}), where the hadron is not on its mass shell. As such three-point functions are not observable, it introduces an uncertainty in the calculation. In order to constrain this uncertainty, it may therefore be worthwhile to provide an alternative evaluation of the HLbL diagram using a dispersive formalism, where the contributing cuts correspond with physical intermediate states. 

As a first step towards this goal, we compute in this paper the massive scalar two-loop three-point functions which are topologically equivalent to the single meson HLbL type of diagrams. 
In fact, the analytic structure of two-loop functions in quantum electro dynamics (QED) is completely defined by the propagators, thus it is identical with the ones of the scalar amplitudes.
Such calculation may subsequently be extended to the calculation of the HLbL contribution to $(g-2)_\m$. 
Indeed, when using monopole form factors for the $M\g^\ast\g^\ast$ vertex in the HLbL tensor the tensor integral appearing in $(g-2)_\m$ may be reduced to the linear combination of scalar integrals of the considered type. The first attempt to calculate two-loop vertex functions of the considered type has been made in \cite{vanNeerven:1985xr}, where the unitarity method was applied to massless diagrams. This was extended in \cite{Korner:1995xd} to the two-loop radiative correction to the Higgs 
boson decay width ($H \to \gamma \gamma$). In this paper we extend the formalism to the case of massive diagrams, allowing for different masses of photons, muons and meson. 

The outline of this paper is as follows. In Section~\ref{sec2}, we set up the formalism to evaluate the two-loop scalar vertex functions, by 
calculating their different absorptive parts. We subsequently discuss the cases of two-particle cuts and three-particle cuts which contribute to the 
absorptive parts. Subsequently, in Section~\ref{sec3}, we test this formalism by comparing the contributions of the different cuts with the direct two-loop calculations for different values of the mass parameters entering the two-loop vertex function. We present our conclusions and outlook in Section~\ref{sec4}. Some technical details which involve the evaluation of the phase space integrals entering the absorptive parts in the dispersion integrals and the direct two-loop calculation are given in two appendices.

\section{Two-loop scalar vertex functions: absorptive part}
\label{sec2}

\begin{figure}[h]
\centering
  \includegraphics[width=11cm]{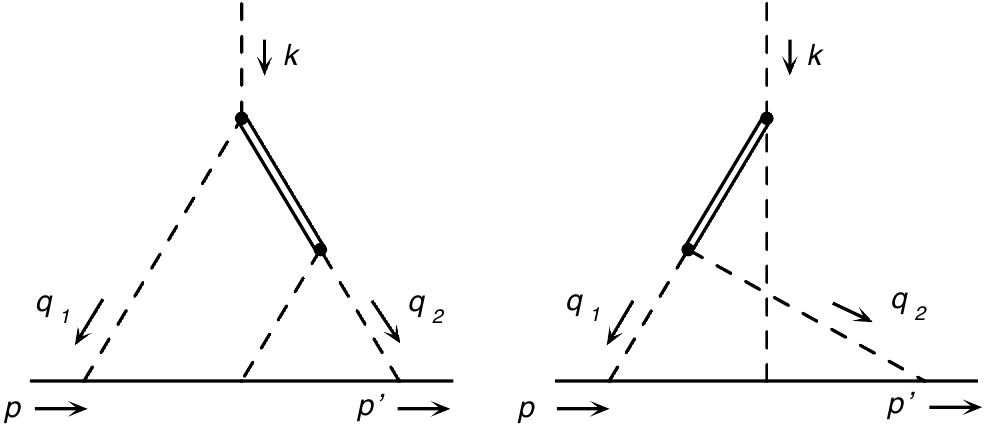}\\
  \caption{Two-loop vertex corrections in a scalar theory. We use the following code for the masses of the scalar propagators: the double line denotes the (meson) propagator with mass $M$, the solid lines denote (lepton) propagators with masses $m_1$ and $m_2$ and the dashed lines denote (photon) propagators with masses $\L_1$, $\L_3$, and $\L_2$ (from left to right).}
  \label{fig:sc2loop}
\end{figure}

For the scalar two-loop amplitude corresponding with the first diagram in Fig. \ref{fig:sc2loop} the Feynman integral reads as
\beq
\begin{split}
\G_{1}(t)=&\int\frac{\dd^4q_1}{(2\p)^4}\int\frac{\dd^4q_2}{(2\p)^4}
\frac{1}{(k-q_1-q_2)^2-\L_3^2}
\frac{1}{q_1^2-\L_1^2}\frac{1}{q_2^2-\L_2^2}\frac1{(k-q_1)^2-M^2}\\
&\hspace{2.5cm}\times\frac{1}{(p+q_1)^2-m_1^2}\frac{1}{(p+k-q_2)^2-m_2^2},
\label{eq:sc2loop1}
\end{split}
\eq
while for the second diagram it has form
\beq
\begin{split}
\G_{2}(t)=&\int\frac{\dd^4q_1}{(2\p)^4}\int\frac{\dd^4q_2}{(2\p)^4}
\frac{1}{(k-q_1-q_2)^2-\L_3^2}
\frac{1}{q_1^2-\L_1^2}\frac{1}{q_2^2-\L_2^2}\frac1{(q_1+q_2)^2-M^2}\\
&\hspace{2.5cm}\times\frac{1}{(p+q_1)^2-m_1^2}\frac{1}{(p+k-q_2)^2-m_2^2}.
\label{eq:sc2loop2}
\end{split}
\eq

Applying the Cutkosky rules to the graphs in Fig. \ref{fig:sc2loop} we can now calculate the absorptive parts of the corresponding amplitudes. 
The absorptive parts can be divided into two-particle cut graphs, shown in Fig.~\ref{fig:sc2pcut}, 
and three-particle cut graphs, shown in Fig.~\ref{fig:sc3pcut}. We will subsequently discuss both cases. 

\subsection{Two-particle cuts} 
\begin{figure}[h]
\centering
  \includegraphics[width=10cm]{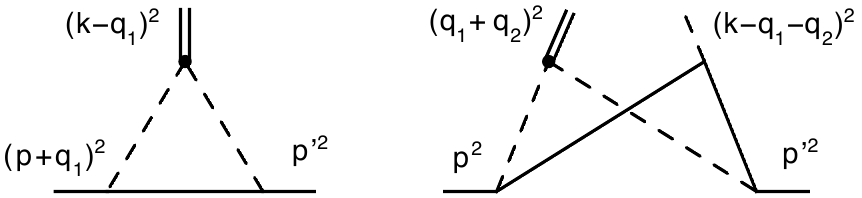}\\
  \caption{One-loop three- and four-point functions in a scalar theory.}
  \label{fig:1looprespfunc}
\end{figure}

The two-particle cut graphs contain the one-loop virtual diagram insertions which are represented by the triangle and box graphs given in Fig. \ref{fig:1looprespfunc}. 
\begin{figure}[h]
\centering
  \includegraphics[width=5cm]{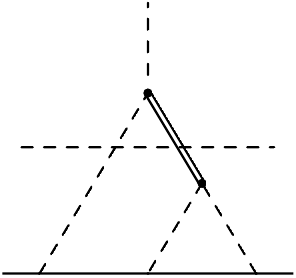}\hspace{1.5cm}
  \includegraphics[width=5cm]{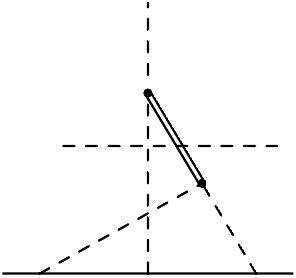}\\
  \caption{The two-particle cut contributions to the two-loop diagrams of Fig. \ref{fig:sc2loop}.}
  \label{fig:sc2pcut}
\end{figure}
With the right choice of the loop-momenta the triangle-diagram correction in the first diagram and the box-diagram correction in the second diagram can be isolated as closed integrals.  The absorptive part corresponding to the left two-particle cut graph in Fig. \ref{fig:sc2pcut} can be expressed in the form:
\beq
\begin{split}
\mathrm{Disc^{2}_t}\G_1(t)&=\int\frac{\dd^4q_1}{(2\p)^4}(-2\p i)^2\d(q_1^2-\L_1^2)\d((k-q_1)^2-M^2)\\
&\hspace{2cm}\times\frac{1}{(p+q_1)^2-m_1^2}M_3\((k-q_1)^2,(p+q_1)^2,m^2,\L_2^2,\L_3^2,m^2\).
\label{eq:12cut}
\end{split}
\eq
To isolate the one-loop function in a closed form in the right two-particle cut diagram in Fig. \ref{fig:sc2pcut} we make a change of variables $q_1\leftrightarrow k-q_1-q_2$. Hence, the discontinuity takes the form:
\beq
\begin{split}
\mathrm{Disc^2_t}\G_2(t)&=\int\frac{\dd^4q_1}{(2\p)^4}(-2\p i)^2\d((q_1^2-\L_3^2)\d((k-q_1)^2-M^2)\\
&\hspace{0cm}\times M_4\((k-q_1)^2,p^2,q_1^2,(p+k)^2,(p+k-q_1)^2,(p+q_1)^2,\L_2^2,\L_1^2,m^2,m^2\).
\label{eq:22cut}
\end{split}
\eq

The one-loop three- and four-point functions $M_3$ and $M_4$ (see Fig. \ref{fig:1looprespfunc}) which enter the expressions for the above loop integrals are given by:
\begin{align}
&M_3\((k-q_1)^2,(p+q_1)^2,m^2,\L_2^2,\L_3^2,m^2\)=\nn\\
&\hspace{4cm}\int\frac{\dd^4q_2}{(2\p)^4}\frac{1}{q_2^2-\L_2^2}\frac{1}{(k-q_1-q_2)^2-\L_3^2}\frac{1}{(p+k-q_2)^2-m^2},
\label{eq:3pointfunc}\\
&M_4\((k-q_1)^2,p^2,q_1^2,(p+k)^2,(p+k-q_1)^2,(p+q_1)^2,\L_2^2,\L_1^2,m^2,m^2\)=\nn\\
&\int\frac{\dd^4q_2}{(2\p)^4}\frac{1}{q_2^2-\L_2^2}\frac{1}{(k-q_1-q_2)^2-\L_1^2}\frac{1}{(p+k-q_1-q_2)^2-m_1^2}\frac{1}{(p+k-q_2)^2-m_2^2}.
\label{eq:4pointfunc}
\end{align}

The above integrals belong to a class of one-loop integrals which were studied in quite detail starting with the original work of Ref. \cite{'tHooft:1978xw} and subsequently extended in Ref. \cite{Denner:2010tr}. The original approach is based on the Feynman parametrization which allows to regroup the propagators in a spherically symmetric form and to perform the momentum integrals in Euclidean space directly. The subsequent integration over Feynman parameters can be performed with the account of the analytical structure of the amplitudes. In regions of momentum space where no cuts occur the integrals are rather simple to perform. In principle, the rest can be obtained by analytical continuation, however in practice it is hard to realize. Therefore, it is more efficient to define the integral independently for different kinematical regions. The details of this approach are described in detail in Refs. \cite{'tHooft:1978xw,Denner:2010tr}.
%It turns out that in particular kinematic region for the case of real masses, the four-point function may be defined in terms of combination of the three-point functions. In the other regimes such a reduction is possible with certain modifications, namely the extra logarithms has to be added.

A big disadvantage of this approach is a very narrow type of parametrizations for the non-perturbative functions which limit its application to rational function parametrizations. It turns out that 
a more efficient way to proceed in this case is by using the dispersion representation of the one-loop integrals of Fig.~\ref{fig:1looprespfunc}. In particular, it becomes useful when considering the $(g-2)_\m$ since its application is not limited to the simplest pole parametrizations of the $\g^\ast \g^\ast \to M$ form factors \cite{paukvdhdisp}. 

Using the angular coordinates and invariants, defined in Appendix~\ref{sec:angularparametrization}, 
the two- and three-particle cuts may be reformulated in a compact form. 
For the two-particle cuts we have
\beq
\begin{split}
\Disc^2_t\G_1(t)&=-\frac{1}{8\p}\int\dd\cos\theta_1\frac{\b_1M_3(M^2,s_-,m^2,\L_2^2,\L_3^2,m^2)}{q_1^2+t_1-t-t\b_1\b\cos\theta_1} ,
\label{eq:2bodyinva}
\end{split}
\eq
and 
\beq
\begin{split}
\Disc^2_t\G_2(t)&=-\frac{1}{16\p}\int\dd\cos\theta_1\;\b_1M_4\(M^2,m^2,\L_3^2,m^2,s_+,s_-,t,\L_2^2,\L_1^2,m^2,m^2\).
\label{eq:2bodyinvb}
\end{split}
\eq
where $\beta_1$ is given by Eq.~(\ref{eqb1}), $s_\pm=m^2+q_1^2/2+t_1/2-t/2 \pm (t/2) \b_1\b\cos\theta_1$, and $t_1 = (k - q_1)^2 = M^2$. 
Note that in Eq.~(\ref{eq:2bodyinva}) $q_1^2=\L_1^2$, whereas in  Eq.~(\ref{eq:2bodyinvb}) $q_1^2=\L_3^2$.

\subsection{Three-particle cuts}

\begin{figure}[h]
\centering
  \includegraphics[width=5cm]{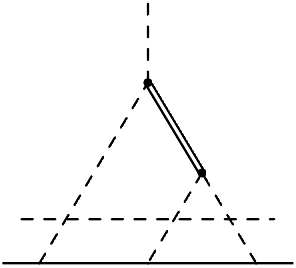}
  \hspace{1cm}
  \includegraphics[width=5cm]{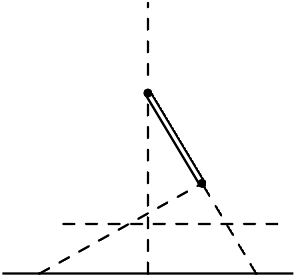}
  \hspace{1cm}
  \includegraphics[width=5cm]{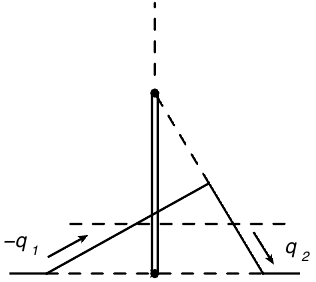}
  \caption{The three-particle cut contributions to the two-loop diagrams of Fig. \ref{fig:sc2loop}. For the third cut diagram (corresponding with the "lepton-lepton-meson" cut) the momenta are relabeled as shown in the figure.}
  \label{fig:sc3pcut}
\end{figure}

The three-particle cut diagrams originating from the two diagrams in Fig. \ref{fig:sc2loop} are depicted in Fig. \ref{fig:sc3pcut}.
For the first three-particle cut in Fig. \ref{fig:sc3pcut} we obtain~:
\beq
\begin{split}
\mathrm{Disc^3_t}\G_1(t)=&\int\frac{\dd^4q_1}{(2\p)^4}\int\frac{\dd^4q_2}{(2\p)^4}(-2\p i)^3\d(q_1^2-\L_1^2)\d(q_2^2-\L_2^2)\d((k-q_1-q_2)^2-\L_3^2)\\
&\hspace{2cm}\times
\frac1{(k-q_1)^2-M^2}\frac{1}{(p+q_1)^2-m_1^2}\frac{1}{(p+k-q_2)^2-m_2^2}.
\label{eq:13cut}
\end{split}
\eq

It occurs that for the second diagram in Fig. \ref{fig:sc2loop}, there are two three-particle cut diagrams which contribute to the discontinuity (see second and third panels in Fig. \ref{fig:sc3pcut}). The first (three-photon) cut contribution (second panel in Fig. \ref{fig:sc3pcut}) is given by
\beq
\begin{split}
\mathrm{Disc^{3,1}_t}\G_2(t)=&\int\frac{\dd^4q_1}{(2\p)^4}\int\frac{\dd^4q_2}{(2\p)^4}(-2\p i)^3\d(q_1^2-\L_1^2)\d(q_2^2-\L_2^2)\d((k-q_1-q_2)^2-\L_3^2)\\
&\hspace{1.5cm}\times
\frac1{(q_1+q_2)^2-M^2}\frac{1}{(p+q_1)^2-m_1^2}\frac{1}{(p+k-q_2)^2-m_2^2},
\label{eq:231cut}
\end{split}
\eq
whereas the second (meson-lepton-lepton) cut contribution (third panel in Fig. \ref{fig:sc3pcut}) is given by
\beq
\begin{split}
\mathrm{Disc^{3,2}_t}\G_2(t)=&\int\frac{\dd^4q_1}{(2\p)^4}\int\frac{\dd^4q_2}{(2\p)^4}(-2\p i)^3\d(q_1^2-m_1^2)\d(q_2^2-m_2^2)\d((k-q_1-q_2)^2-M^2)\\
&\hspace{2cm}\times\frac1{(q_1+q_2)^2-\L_3^2}\frac{1}{(p+q_1)^2-\L_1^2}\frac{1}{(p+k-q_2)^2-\L_2^2}.
\label{eq:232cut}
\end{split} 
\eq

\begin{figure}[h]
\centering
  \includegraphics[width=7cm]{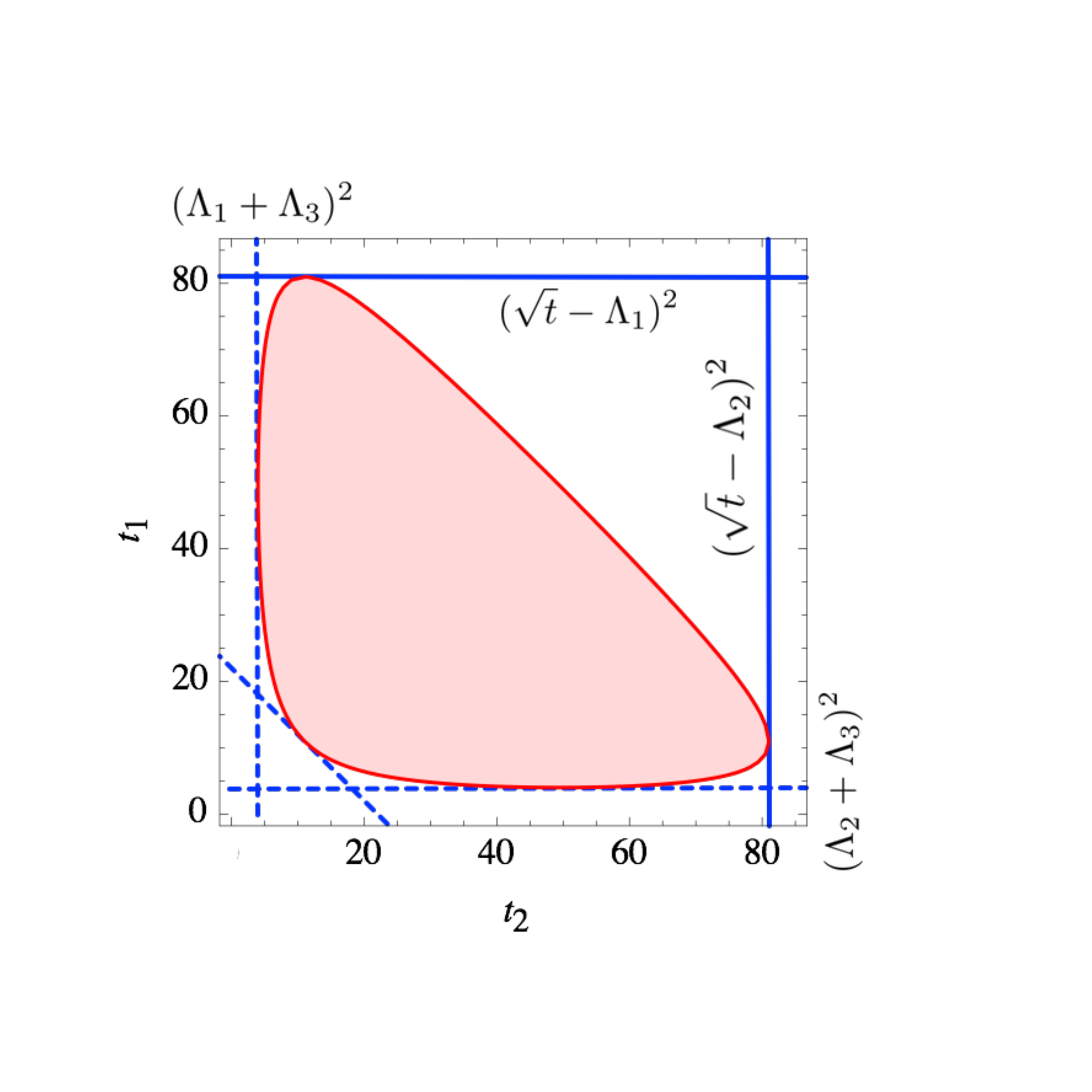}\\
  \caption{The integration domain for the three-particle cut. The blue dashed lines correspond to the conditions in Eq. (\ref{eq:3pintcondb}); the solid blue lines correspond to Eqs. (\ref{eq:3pintconds1}, \ref{eq:3pintconds2}); the red curve represents the 
  condition $-1\leqslant\cos\theta\leqslant1$, with $\cos \theta$ as defined in Eq.~(\ref{eq:q1q2_kinematics}).}
  \label{fig:3pintegration}
\end{figure}

The phase-space integrals in the discontinuities of Eqs. (\ref{eq:13cut}-\ref{eq:232cut}) may be partly performed analytically by introducing the angular parametrization, as described in  
Appendix \ref{sec:angularparametrization}.   
For the three-particle discontinuity in the first diagram, defined by Eq. (\ref{eq:13cut}), this yields~: 
\beq
\begin{split}
\mathrm{Disc_t}\G_1^{(3)}(t)=&i\frac1{2(4\p)^4t}\int\dd t_1\int\dd t_2\frac1{t_1-M^2}\;\\
&\hspace{2cm}\times\Omega(t_1,t_2,t+\L_1^2+\L_2^2+\L_3^2-t_1-t_2,\L_1^2,\L_2^2,m_1^2,m_2^2).
\label{eq:sc2loop13pcut1}
\end{split}
\eq
For the discontinuities coming from the second three-particle cut diagram in Fig. \ref{fig:sc3pcut} defined by Eq. (\ref{eq:231cut}) we 
obtain~:
\beq
\begin{split}
\mathrm{Disc_t}\G_2^{(3,1)}(t)=&i\frac1{2(4\p)^4t}\int\dd t_1\int\dd t_2\frac1{t+\L_1^2+\L_2^2+\L_3^2-t_1-t_2-M^2}\\
&\hspace{2cm}\times\Omega(t_1,t_2,t+\L_1^2+\L_2^2+\L_3^2-t_1-t_2,\L_1^2,\L_2^2,m_1^2,m_2^2).
\label{eq:sc2loop13pcut21}
\end{split}
\eq
The integral over the variables $(t_1, t_2)$ is performed over a two-dimensional region shown in Fig. \ref{fig:3pintegration}. It is defined by the inequalities:
\begin{align}
q_1^0&\geqslant\L_1\;&\Rightarrow\;&&t_1&\leqslant(\sqrt{t}-\L_1)^2\label{eq:3pintconds1},\\
q_2^0&\geqslant\L_2\;&\Rightarrow\;&&t_2&\leqslant(\sqrt{t}-\L_2)^2\label{eq:3pintconds2},\\
k^0-q_1^0-q_2^0&\geqslant\L_3\;&\Rightarrow\;&&t_1+t_2&\geqslant2\sqrt{t}\L_3+\L_1^2+\L_2^2,
\end{align}
and
\beq
t_1\geqslant(\L_3+\L_2)^2,\qquad t_2\geqslant(\L_1+\L_3)^2.
\label{eq:3pintcondb}
\eq
An additional constraint is imposed by the condition $-1\leqslant\cos\theta\leqslant1$ which results in
\beq
-1\leqslant\frac{2t(q_1^2+q_2^2-t_{12})+(t-t_1+q_1^2)(t-t_2+q_2^2)}{t^2\b_1\b_2}\leqslant1.
\eq

For the discontinuities coming from the third three-particle cut diagram in Fig. \ref{fig:sc3pcut} the analogous expressions may be easily obtained by substitutions $m_1\leftrightarrow\L_1$, $m_2\leftrightarrow\L_2$, $M\leftrightarrow\L_3$.  As a result, 
we obtain the expression for the cut in Eq. (\ref{eq:232cut}):
\beq
\begin{split}
\mathrm{Disc_t}\G_2^{(3,2)}(t)=&i\frac1{2(4\p)^4t}\int\dd t_1\int\dd t_2\frac1{t+m_1^2+m_2^2+M^2-t_1-t_2-\L_3^2}\\
&\hspace{1cm}\times\Omega(t_1,t_2,t+m_1^2+m_2^2+M^2-t_1-t_2,m_1^2,m_2^2,\L_1^2,\L_2^2).
\label{eq:sc2loop13pcut22}
\end{split}
\eq

\section{Results and discussion}
\label{sec3}

\begin{figure}[h]
\centering
  \includegraphics[width=7.4cm]{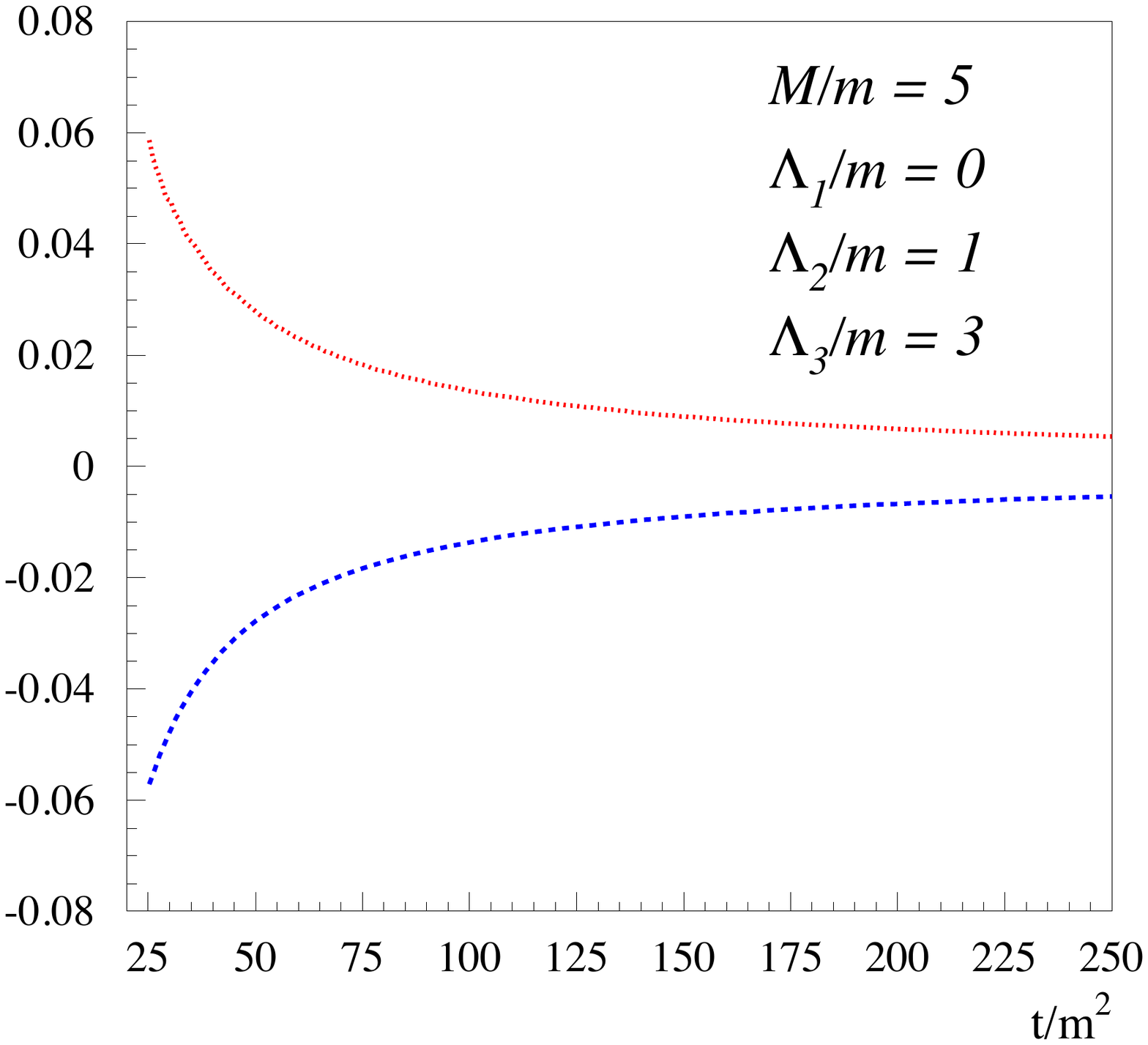}
  \includegraphics[width=7.4cm]{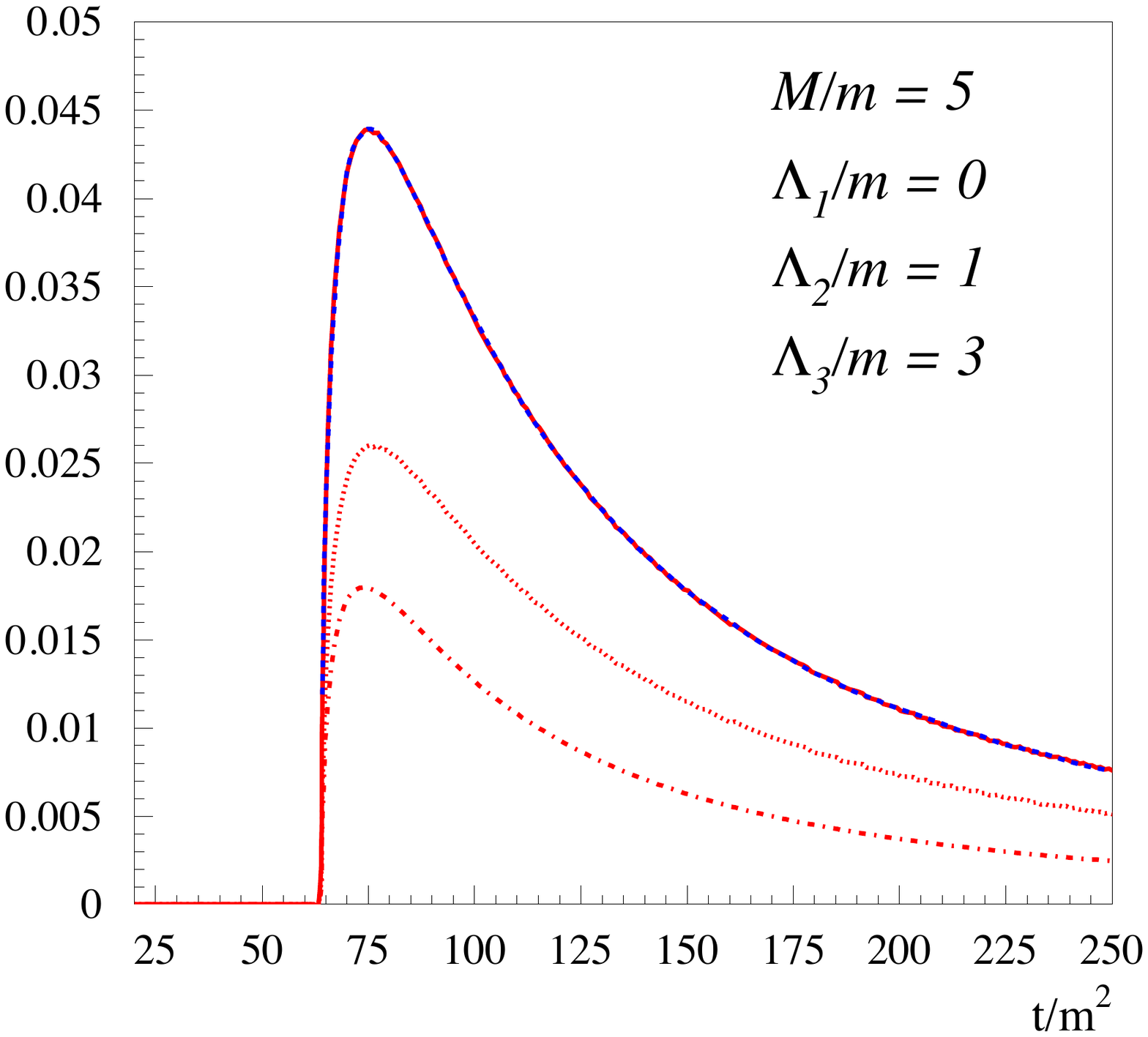}
  \vspace{-2cm}
  \caption{The absorptive parts of the vertex function discontinuities. 
  Left panel: the absorptive parts of the two- and three-particle discontinuities of the first diagram. The red dotted (blue dashed) curve denotes the absorptive part of the three-particle discontinuity (two-particle discontinuity).
  Right panel: the absorptive parts of the two- and three-particle discontinuities of the second diagram. The dotted and dash-dotted red curves denote the absorptive parts of the three-particle discontinuities (three "photon" cut and "lepton-lepton-meson" cut respectively), and the solid red curve denotes the sum of both. The blue dashed curve stands for the absorptive part of the two-particle discontinuity (shown with the opposite sign to demonstrate the cancellation).}
  \label{fig:impart_discontinuities}
\end{figure}

We will start from a discussion of the absorptive parts of the two- and three-particle discontinuities. The sum of both should vanish as the full discontinuity has to be a purely imaginary function. The absorptive parts may be easily obtained by re-applying Cutkosky rules to the cut diagrams in Figs. \ref{fig:sc2pcut} and \ref{fig:sc3pcut}. For the first diagram the absorptive part of the three-particle discontinuity originates from the propagator $(t_1-M^2+i\ve)^{-1}$; for the second diagram from propagators $(t_{12}-M^2+i\ve)^{-1}$ and $(t_{12}-\L_3^2+i\ve)^{-1}$ for the first and the second cut respectively. The absorptive part of the two-particle discontinuity is coming from the two-particle cut of the triangle and box diagrams in 
Fig.~\ref{fig:1looprespfunc}. For the sake of demonstration we will consider the mass configuration $\L_1=0$, $\L_2=m$, $\L_3=3m$ and $M=5m$. The absorptive parts of the discontinuities are shown in Fig.~\ref{fig:impart_discontinuities}. We can see that the absorptive part starts exactly at the two-particle (meson-photon) thresholds $(\L_1+M)^2$ for the first diagram and $(\L_3+M)^2$ for the second. As is expected, for both topologies the absorptive parts of the two- and three-particle cuts are identical and have opposite signs, such that the discontinuity is a pure imaginary function. 
\vspace{-1.5cm}

\begin{figure}[h]
\centering
  \includegraphics[width=7.4cm]{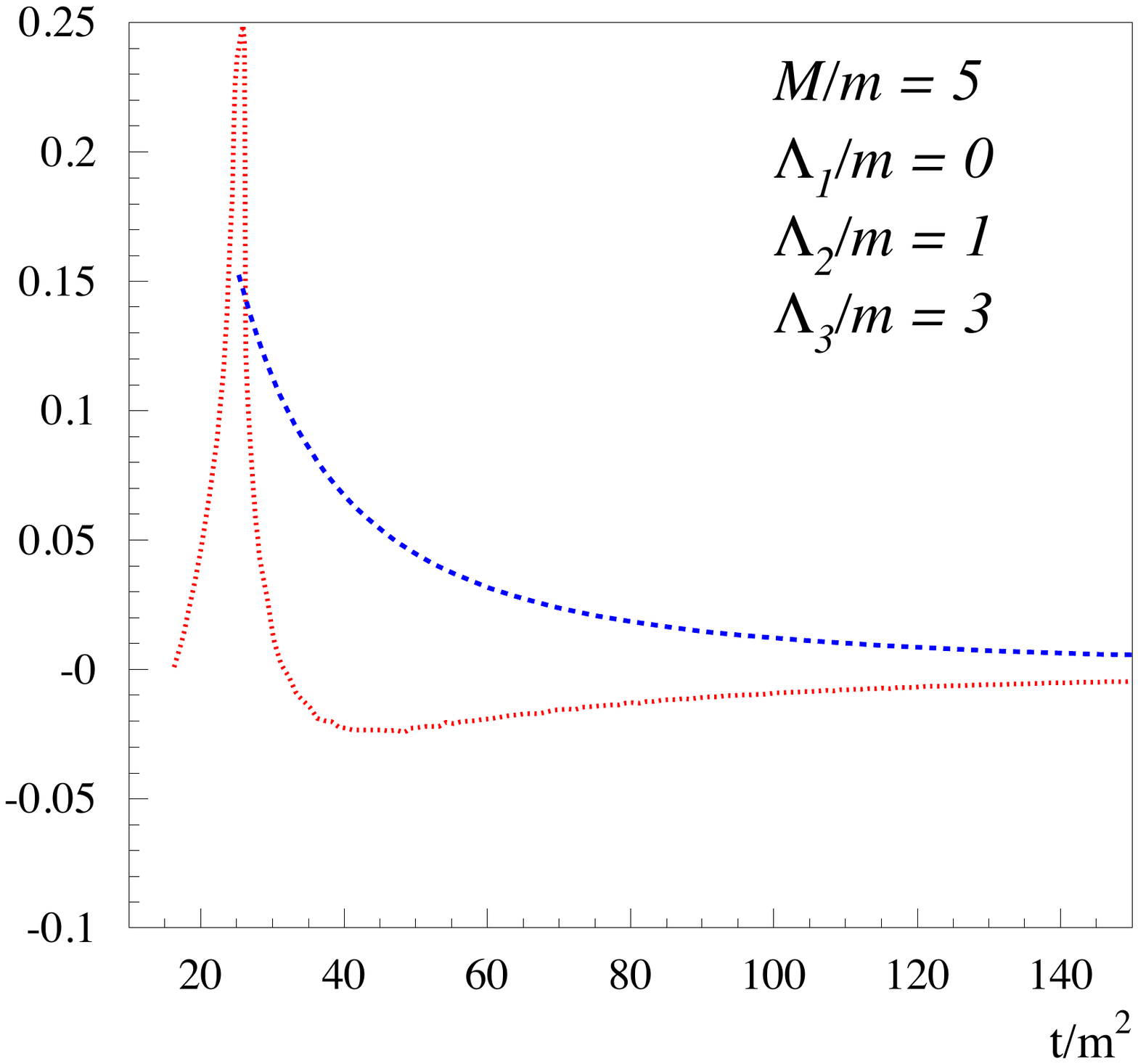}
  \includegraphics[width=7.4cm]{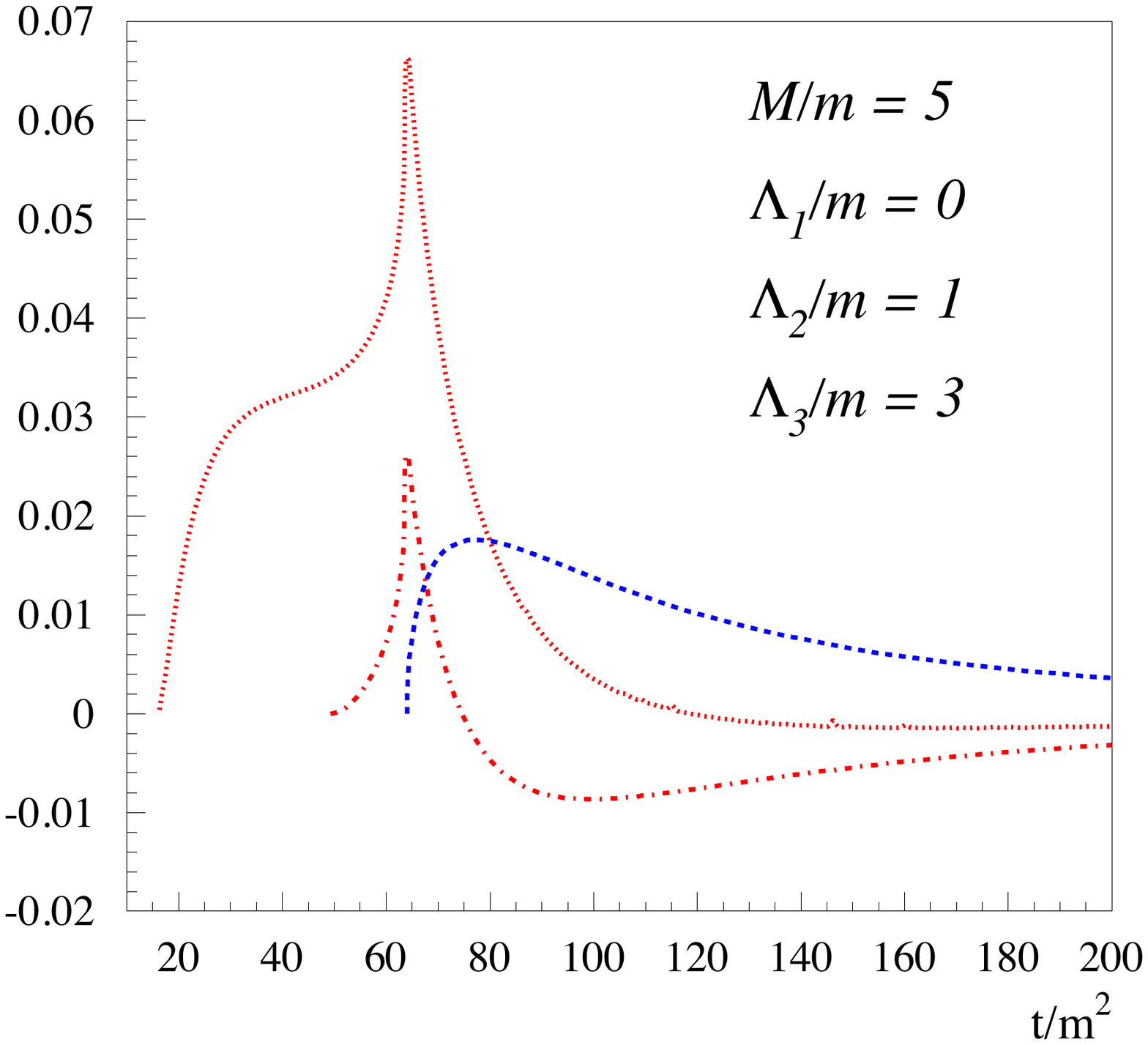}
  \vspace{-2cm}
  \caption{The real parts of the vertex function discontinuities: Im $\G_i(t)/t$. 
  Left panel: the real parts of the two- and three-particle discontinuities of the first diagram. The red dotted (blue dashed) curve denotes the real part of the three-particle (two-particle) discontinuity.  Right panel: the real parts of the two- and three-particle discontinuities of the second diagram. The red dotted and dash-dotted curves denote the real part of the three-particle discontinuities originating from the 
 first (three "photon") cut and second   
("lepton-lepton-meson") cut respectively. The blue dashed curve stands for the real part of the two-particle discontinuity.}
  \label{fig:impartscalar}
\end{figure}

The $t$-dependence of the imaginary parts of the vertex functions (real part of the discontinuities), defined as 
${\mathrm {Im} \Gamma (t)} \equiv 1/(2 i) \mathrm{Disc}_t \Gamma(t)$, is shown on the plots in Fig. \ref{fig:impartscalar}. We can clearly see that these discontinuities start at the corresponding two- and three-particle thresholds. In the case of the first diagram, it is $(\L_1+M)^2$ for the two-particle cut and $(\L_1+\L_2+\L_3)^2$ for the three-particle cut. Regarding the second diagram, the two-particle cut starts at $(\L_3+M)^2$ and the three-particle cut starts  at $(\L_1+\L_2+\L_3)^2$ and $(M+2m)^2$ for the first (three "photon") cut and the second ("meson-lepton-lepton") cut respectively. We can observe clear cusps in the three-particle cuts exactly at the position where the two-particle cut starts. It corresponds to the opening of a new, two-particle ("meson-photon") channel and is correlated with the threshold shown in Fig.~\ref{fig:impart_discontinuities}.
\vspace{-1.5cm}

The real part of the vertex function $\Gamma(0)$ can now be obtained directly by performing the dispersion integral over $t$ as~:
\begin{align}
\G(0)&=\frac1{2\p i}\int\limits_{(\L_1+M)^2}^\infty\frac {\dd t}{t}\;\mathrm{Disc^{(2)}_t}\G_1(t)+\frac1{2\p i}\int\limits_{(\L_3+M)^2}^\infty\frac {\dd t}{t}\;\mathrm{Disc^{(2)}_t}\G_2(t)\\
&+\frac1{2\p i}\int\limits_{(\L_1+\L_2+\L_3)^2}^\infty\frac {\dd t}{t}\;\[\mathrm{Disc^{(3)}_t}\G_1(t)+\mathrm{Disc^{(3,1)}_t}\G_2(t)\]
+\frac1{2\p i}\int\limits_{(2m+M)^2}^\infty\frac {\dd t}{t}\;\mathrm{Disc^{(3,2)}_t}\G_2(t).\nn
\end{align}

\begin{figure}[h]
\centering
  \includegraphics[width=7.4cm]{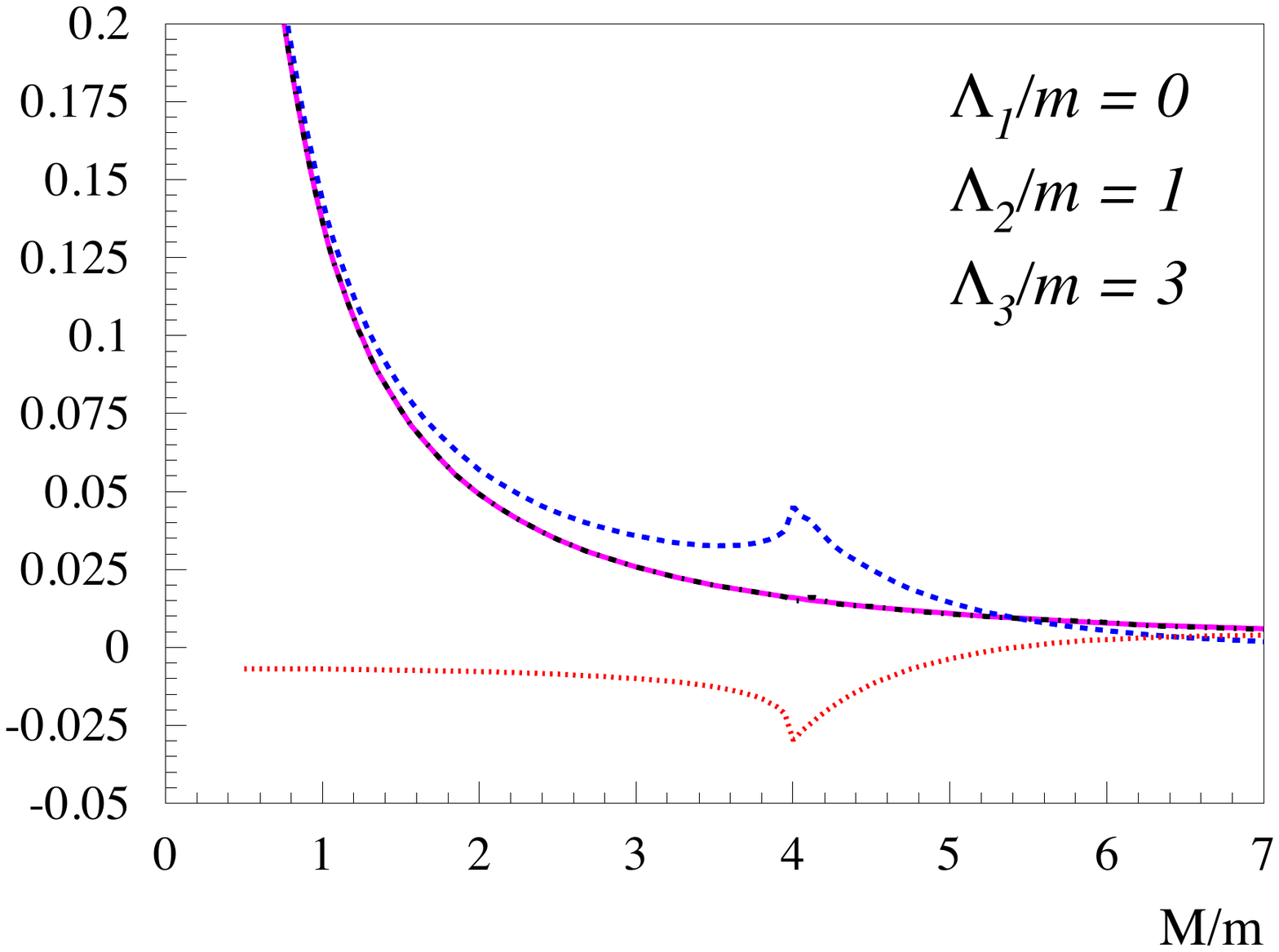}
  \includegraphics[width=7.4cm]{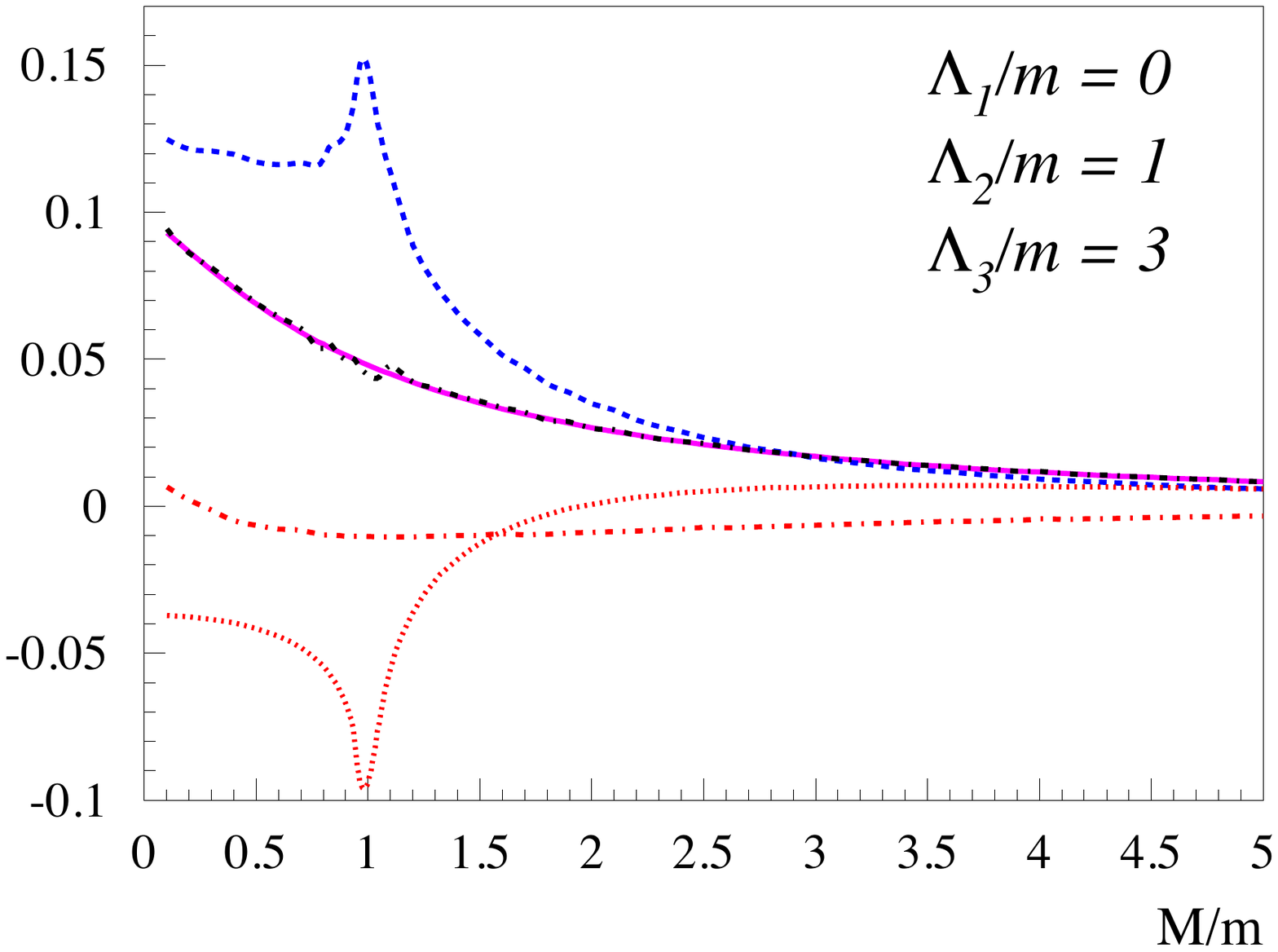}
  \vspace{-2cm}
  \caption{The dependence of the vertex function at zero momentum transfer, $\Gamma(0)$, on the (meson) mass $M$. 
  Left panel: the contributions which originate from the two- and three-particle discontinuities of the first diagram. The red dotted (blue dashed) curve denotes the contribution of the three-particle (two-particle) discontinuity. The black dashed-dotted curve denotes the sum of both contributions. The pink solid curve is obtained by the direct evaluation of two-loop integrals.
  Right panel: the contributions of the two- and three-particle discontinuities of the second diagram. The dotted and dash-dotted red curves denote the contributions of the three-particle discontinuities (three "photon" cut and "lepton-lepton-meson" cut  respectively). The blue dashed curve stands for the contribution of the two-particle discontinuity. The sum of all three contributions is given by black dashed-dotted curve. The pink solid curve is obtained by the direct evaluation of  two-loop integrals.}
  \label{fig:vertexdepM}
\end{figure}

The dependence of $\Gamma(0)$, representing the form factor at zero momentum transfer, on the (meson) mass $M$ is shown in Fig. \ref{fig:vertexdepM}. We can observe spikes in both two- and three-particle cuts located at $M=\L_2+\L_3$ for the first diagram and $M=\L_1+\L_2$ for the second. This cusp can be attributed to the opening of the two-particle threshold when considering the one-loop diagrams depending on the virtuality $t_1$. Since in the dispersion evaluation we put the meson on its mass shell, i.e. $t_1=M^2$ this effect is reflected in the dependencies of the contribution of two- and three-particle discontinuities on $M$. When performing the loop integral directly, the meson is not an external particle but rather virtual and this effect does not emerge. Practically, when applied to the anomalous magnetic moment calculation it means that for light hadronic states or for the states with mass $M\sim \L$ with $\L$ a monopole mass parameter of the form factor parametrization, we need additional precision in the data input. We notice from Fig. \ref{fig:vertexdepM} that the sum of all 
cut contributions exactly reproduces the dependence obtained by the direct evaluation of the loop integrals using the hyperspherical approach (see Appendix \ref{app:2loopgeg}). 

We thus provided an explicit demonstration 
that the developed dispersion technique can be applied to the calculation of the HLbL correction to $(g-2)_\m$. 
\vspace{-2cm}

\section{Conclusion}
\label{sec4}

In the present work, we have presented a dispersion relation formalism to calculate a scalar two-loop vertex function where the internal lines have different masses. The considered topologies are of relevance for the calculation of the hadronic light-by-light (HLbL) contribution to the muon's anomalous magnetic moment when 
including the internal structure of a meson in the loop through form factors. We have shown that in the dispersive formalism the absorptive part of the two-loop diagram is obtained by a sum of two- and three-particle cut contributions. 
Each of these cut contributions involves a phase space integration over the physical intermediate states. 
Using the absorptive parts as input, we have reconstructed the real part of the vertex through a dispersion relation. We have explicitly demonstrated for the case of a scalar vertex, with different masses, that when summing all cut contributions, the dispersive formalism yields exactly the same result as the direct two-loop calculation. As the intermediate states in the dispersive formalism 
are on their mass-shell, such an evaluation avoids any off-shell ambiguities which may arise when evaluating the two-loop diagram with composite vertices. Therefore, the present formalism may serve as 
the basis to extend such calculation to the case of the HLbL contribution to the muon's anomalous magnetic moment, when considering meson exchange with form factors. Such a calculation will be presented in a future work.

\section*{Acknowledgements}

This work was supported by the Deutsche Forschungsgemeinschaft DFG in part through the Collaborative 
Research Center ``The Low-Energy Frontier of the Standard Model" (SFB 1044), 
in part through the graduate school Graduate School ``Symmetry Breaking in Fundamental Interactions" 
(DFG/GRK 1581), and in part through 
the Cluster of Excellence "Precision Physics, Fundamental Interactions and Structure of Matter" (PRISMA).

%\clearpage
\appendix

\section{Angular parametrization and integration of the phase space}
\label{sec:angularparametrization}

The scalar two-loop integrals can be represented in a convenient way using a special set of variables. The choice of variables is based on the initial form of the integrand: the non-perturbative vacuum polarization part is dependent only on invariants and the angular integral is reduced to the integration over one polar and azimuthal angle. We meet the same type of the angular integral when integrating the three-particle phase space. It can be evaluated analytically and expressed in a closed form. In this section we discuss the technical details of the angular integration.% and the invariants' phase space appearing in the integrals for the discontinuities.

\begin{figure}[h]
\centering
  \includegraphics[width=10cm]{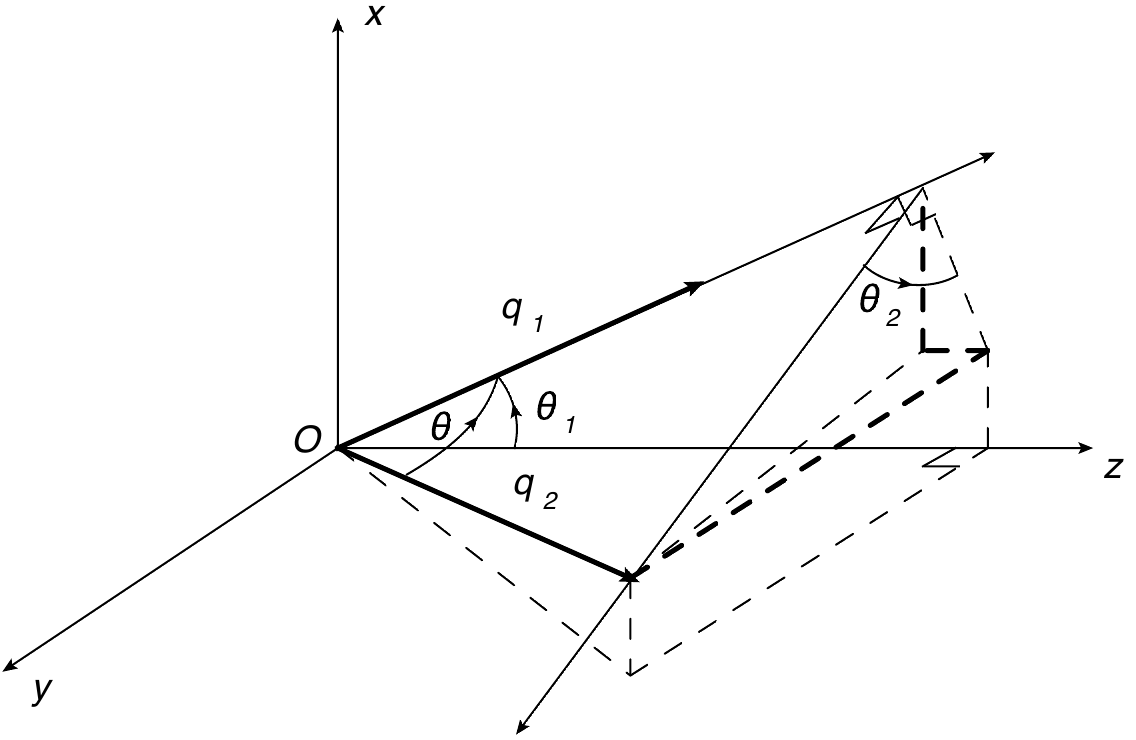}\\
  \caption{The angular coordinates in the one- and two-loop phase space integral.}
  \label{3dimPhSp_angular_integration}
\end{figure}

The most important ingredient of the dispersion method is the calculation of the angular part of the two- and three-particle phase-space integrals. The loop and external momenta can be defined in terms of the angular coordinates, a set of invariant energy parameters $t$, $t_1$, $t_2$, $t_{12}$, 
and the invariant masses of virtual photons $q_1^2$ and $q_2^2$. In the $k$-rest frame we define the momenta as
\beq
\begin{split}
k&=\(\sqrt{t},0,0,0\),\\
p&=\frac{\sqrt{t}}2\(-1,0,0,\b\),\\
q_1&=\frac{\sqrt{t}}2\b_1\(\frac{t-t_1+q_1^2}{t\b_1},\sin\theta_1,0,\cos\theta_1\),\\
q_2&=\frac{\sqrt{t}}2\b_2\(\frac{t-t_2+q_2^2}{t\b_2},-\cos\theta_1\cos\theta_2\sin\theta+\sin\theta_1\cos\theta,\right.\\
&\hspace{2cm}\left.\sin\theta\sin\theta_2,\sin\theta_1\cos\theta_2\sin\theta+\cos\theta_1\cos\theta\),\\
\cos\theta&=\frac{2t(q_1^2+q_2^2-t_{12})+(t-t_1+q_1^2)(t-t_2+q_2^2)}{t^2\b_1\b_2}.
\end{split}
\label{eq:q1q2_kinematics}
\eq
with
\beq
\b_i=\sqrt{\(1+\frac{q_i^2-t_i}{t}\)^2-\frac{4q_i^2}{t}}\quad\mathrm{and}\quad\b=\sqrt{1-\frac{4m^2}{t}}.
\label{eqb1}
\eq
This angular parametrization is explained by Fig.(\ref{3dimPhSp_angular_integration}). The space direction of the momentum $q_2$ is defined with respect to $q_1$, by a polar angle $\theta$ between $\vec{q}_1$ and $\vec{q}_2$ and azimuthal angle $\theta_2$. Such a definition allows to factorize the two-loop expression and to express the angular integral in a closed form.

The invariants which appear in the calculation are related to the introduced parameters as

\beq
\begin{split}
p^2=p'^2&=m^2,\\
(k-q_1)^2&=t_1,\\
(k-q_2)^2&=t_2,\\
(q_1+q_2)^2&=t_{12},\\
(p+q_1)^2&=\frac12\[2m^2+q_1^2+t_1-t-t\b_1\b\cos\theta_1\],\\
(p+k-q_1)^2&=\frac12\[2m^2+q_1^2+t_1-t+t\b_1\b\cos\theta_1\],\\
(p+k-q_2)^2&=\frac12\[2m^2+q_2^2-t+t_2+t\b_2\b(\sin\theta_1\cos\theta_2\sin\theta+\cos\theta_1\cos\theta)\],\\
(k-q_1-q_2)^2&=t_{12}+t_1+t_2-t-q_1^2-q_2^2,\\
\end{split}
\label{eq:q1q2_kinematics2}
\eq

Using the above definitions the transformation of the phase-space integration measure is given by
\begin{align}
\int\dd^4q_1&=\int\limits_{}^{}\dd q_1^2\int\limits_{}^{}\dd t_1 \int\limits_0^{\p}\dd\cos\theta_1\int\dd\phi\;\frac{\b_1}8,\\
\int\dd^4q_2&=\int\dd q_2^2\int\dd t_2\int\dd t_{12}\int\dd\theta_2\;\frac{1}{4\b_1t}
\end{align}
where the integration domains are explicitly defined by kinematic constraints for each of two- and three-particle cuts separately. 
%For two-loop integral\beq\int\dd^4q_1\int\dd^4q_2=\frac{1}{32t}\int\limits_{}^{}\dd q_1^2\int\dd q_2^2\int\limits_{}^{}\dd t_1\int\dd t_2\int\dd t_{12} \int\limits_0^{\p}\dd\cos\theta_1\int\dd\theta_2\int\dd\phi\eq
Using the introduced parametrization the loop integrals in Eqs. (\ref{eq:sc2loop1}, \ref{eq:sc2loop2}) can be represented as
\beq
\begin{split}
\G_{1}(t)=&\frac4{(4\p)^7t}\int\dd q_1^2\int\dd q_2^2\int\dd t_1\int\dd t_2\int\dd t_{12}
\frac{1}{t_{12}+t_1+t_2-t-q_1^2-q_2^2-\L_3^2}
\\
&\hspace{3.5cm}\times\frac{1}{q_1^2-\L_1^2}\frac{1}{q_2^2-\L_2^2}\frac1{t_1-M^2}\Omega(t_1,t_2,t_{12},q_1^2,q_2^2,m_1^2,m_2^2),
\label{eq:sc2loop1ang}
\end{split}
\eq
and
\beq
\begin{split}
\G_{2}(t)=&\frac4{(4\p)^7t}\int\dd q_1^2\int\dd q_2^2\int\dd t_1\int\dd t_2\int\dd t_{12}
\frac{1}{t_{12}+t_1+t_2-t-q_1^2-q_2^2-\L_3^2}
\\
&\hspace{2.5cm}\times\frac{1}{q_1^2-\L_1^2}\frac{1}{q_2^2-\L_2^2}\frac1{t_{12}-M^2}\Omega(t_1,t_2,t_{12},q_1^2,q_2^2,m_1^2,m_2^2) ,
\label{eq:sc2loop2ang}
\end{split}
\eq
where $\Omega(t_1,t_2,t_{12},q_1^2,q_2^2,m_1^2,m_2^2)$ is the angular integral defined by
\beq
\begin{split}
&\Omega(t_1,t_2,t_{12},q_1^2,q_2^2,m_1^2,m_2^2)=\int\limits_0^{\p}\dd\cos\theta_1\int\limits_0^{2\p}\dd\theta_2\;\frac2{2m^2-2m_1^2+q_1^2+t_1-t-t\b_1\b\cos\theta_1}\\
&\hspace{2.4cm}\times\frac2{2m^2-2m_2^2+q_2^2-t+t_2+t\b_2\b(\sin\theta_1\cos\theta_2\sin\theta+\cos\theta_1\cos\theta)}.
\end{split}
\eq
The angular integral $\Omega$ can be carried out analytically which yields  
\beq
\begin{split}
&\Omega(t_1,t_2,t_{12},q_1^2,q_2^2,m_1^2,m_2^2)=-\frac{4}{\b_1\b_2\b^2t^2}\frac{2\p}{\sqrt{(a+b)^2-2ab(1+\cos\theta)-\sin^2\theta}}\\
&\times\[\log\frac{a-1}{a+1}+\log\frac{(a+b)(1-b)-b(1+\cos\theta)+ab(1+\cos\theta)+\sin^2\theta+(\cos\theta+b)c}{-(a+b)(1+b)+b(1+\cos\theta)+ab(1+\cos\theta)+\sin^2\theta-(\cos\theta-b)c}\],
\end{split}
\eq
with
\beq
\begin{split}
a&=-\frac{2m^2-2m_1^2+q_1^2-t+t_1}{t\b_1\b},\\
b&=\frac{2m^2-2m_2^2+q_2^2-t+t_2}{t\b_2\b},\\
c&=\sqrt{(a+b)^2-2ab(1+\cos\theta)-\sin^2\theta}.
\end{split}
\eq

\section{Two-loop scalar vertex function in the hyperspherical approach}
\label{app:2loopgeg}
In this appendix we present the direct evaluation of the two-loop integral entering the scalar vertex. The approach is based on the properties of the Gegenabuer polynomials. From the generating function, we obtain the following representation of the propagators in Euclidean space:
\begin{align}
\frac1{(K-L)^2+M^2}&=\frac{Z^M_{KL}}{|K||L|}\sum\limits_{n=0}^\infty\(Z^M_{KL}\)^nC_n(\hat{K}\cdot\hat{L}),
\label{Chebser1}\\
\frac1{(K+L)^2+M^2}&=\frac{Z^M_{KL}}{|K||L|}\sum\limits_{n=0}^\infty\(-Z^M_{KL}\)^nC_n(\hat{K}\cdot\hat{L}),
\label{Chebser2}\\
Z^M_{KL}&=\frac{K^2+L^2+M^2-\sqrt{(K^2+L^2+M^2)^2-4K^2L^2}}{2|K||L|},
\end{align}
where $C_n(x)$ are the Gegenbauer polynomials. 
The orthogonality relation allows to perform hyperangular integrals analytically:
\begin{align}
\int \dd\Omega (\hat{K}) C_n(\hat{Q}_1\cdot \hat{K}) C_m(\hat{K}\cdot\hat{Q}_2)&=2\p^2\frac{\d_{nm}}{n+1}C_n(\hat{Q}_1\cdot\hat{Q}_2), 
\label{Cheb_orth}\\
\int \dd\Omega (\hat{K}) C_n(\hat{Q}\cdot \hat{K}) C_m(\hat{K}\cdot\hat{Q})&=2\p^2\d_{nm}.
\end{align}

The contribution of the first diagram in Fig.(\ref{fig:sc2loop}) is
\beq
\begin{split}
\G_1(0)=\int\frac{\dd^4q_1}{(2\p)^4}\int\frac{\dd^4q_2}{(2\p)^4}
&\frac{1}{q_1^2-M^2}\frac{1}{q_1^2-\L_1^2}\frac{1}{q_2^2-\L_2^2}\\
\times&\frac{1}{(p+q_1)^2-m^2}\frac{1}{(p-q_2)^2-m^2}\frac{1}{(q_1+q_2)^2-\L_3^2}.
\label{eq:scloop1app}
\end{split}
\eq
After Wick rotation $q_i^0\to iQ_i^0$ the two-loop integral takes the form:
\beq
\begin{split}
\G_1(0)=-\int\frac{\dd^4Q_1}{(2\p)^4}\int\frac{\dd^4Q_2}{(2\p)^4}
&\frac{1}{Q_1^2+M^2}\frac{1}{Q_1^2+\L_1^2}\frac{1}{Q_2^2+\L_2^2}\\
\times&\frac{1}{(P+Q_1)^2+m^2}\frac{1}{(P-Q_2)^2+m^2}\frac{1}{(Q_1+Q_2)^2+\L_3^2}.
\end{split}
\eq
Changing the integral measure to the hyperspherical coordinates we obtain:
\beq
\begin{split}
\G_1(0)=&-\frac{1}{4(2\p)^4}\int\dd Q_1\,Q_1^3\int\dd Q_2\,Q_2^3\int\frac{\dd \Omega(Q_1)}{2\p^2}\int\frac{\dd \Omega(Q_2)}{2\p^2}
\frac{1}{Q_1^2+M^2}\frac{1}{Q_1^2+\L_1^2}\frac{1}{Q_2^2+\L_2^2}\\
\times&\frac{1}{(P+Q_1)^2+m^2}\frac{1}{(P-Q_2)^2+m^2}\frac{1}{(Q_1+Q_2)^2+\L_3^2}.
\end{split}
\eq

Using properties of Gegenbauer polynomials we can express the vertex function of Eq. (\ref{eq:scloop1app}) in terms of 
a two-dimensional integral representation:
\beq
\begin{split}
\G_1(0)=-\frac{1}{4(2\p)^4}&\int\dd Q_1\int\dd Q_2
\;\frac{Q_1^3}{Q_1^2+\L_1^2}\frac{Q_2^3}{Q_2^2+\L_2^2}\frac{1}{Q_1^2+M^2}\\
\times&\frac1{m^2Q_1^2Q_2^2}\ln\[1+\frac{(\L_3^2+Q_1^2+Q_2^2-R^{\L_3})(Q_1^2-R_1^m)(Q_2^2-R_2^m)}{8m^2Q_1^2Q_2^2}\].
\end{split}
\eq

For the second diagram the two-loop integral in the hyperspherical coordinates takes form:
\beq
\begin{split}
\G_2(0)=&-\frac{1}{4(2\p)^4}\int\dd Q_1\int\dd Q_2\int\frac{\dd \Omega(Q_1)}{2\p^2}\int\frac{\dd \Omega(Q_2)}{2\p^2}
Q_1^3Q_2^3\frac{1}{Q_1^2+\L_1^2}\frac{1}{Q_2^2+\L_2^2}\\
\times&\frac{1}{(P+Q_1)^2+m^2}\frac{1}{(P-Q_2)^2+m^2}\frac{1}{(Q_1+Q_2)^2+\L_3^2}\frac{1}{(Q_1+Q_2)^2+M^2}.
\end{split}
\eq
A fractional decomposition of the denominator yields:
\beq
\begin{split}
\G_2(0)&=-\frac{1}{4(2\p)^4}\int\dd Q_1\int\dd Q_2\int\frac{\dd \Omega(Q_1)}{2\p^2}\int\frac{\dd \Omega(Q_2)}{2\p^2}
Q_1^3Q_2^3\frac{1}{Q_1^2+\L_1^2}\frac{1}{Q_2^2+\L_2^2}\\
\times&\frac{1}{(P+Q_1)^2+m^2}\frac{1}{(P-Q_2)^2+m^2}\frac{1}{M^2-\L_3^2}\[\frac{1}{(Q_1+Q_2)^2+\L_3^2}-\frac{1}{(Q_1+Q_2)^2+M^2}\],
\end{split}
\eq
which can be easily integrated using the orthogonality relations of Eq. (\ref{Cheb_orth}) and power expansions of Eq. (\ref{Chebser1}, \ref{Chebser2}). As a result we get:
\beq
\begin{split}
\G_2(0)=-\frac{1}{4(2\p)^4}\int\dd Q_1\int\dd Q_2
&\;\frac{Q_1^3}{Q_1^2+\L_1^2}\frac{Q_2^3}{Q_2^2+\L_2^2}\frac{1}{M^2-\L_3^2}\frac1{m^2Q_1^2Q_2^2}\\
\times&
\[\ln\[1+\frac{(\L_3^2+Q_1^2+Q_2^2-R^{\L_3})(Q_1^2-R_1^m)(Q_2^2-R_2^m)}{8m^2Q_1^2Q_2^2}\]\right.\\
-&\left.\ln\[1+\frac{(M^2+Q_1^2+Q_2^2-R^M)(Q_1^2-R_1^m)(Q_2^2-R_2^m)}{8m^2Q_1^2Q_2^2}\]\].
\label{eq:scverths}
\end{split}
\eq
The obtained two-dimensional representation may be integrated numerically.

\end{document}